*Article*

# 3D large-scale fused silica microfluidic chips enabled by hybrid laser microfabrication for continuous-flow UV photochemical synthesis

Aodong Zhang [1,2,3], Jian Xu [1,2,3,*], Yucen Li [1], Ming Hu [1], Zijie Lin [2,3], Yunpeng Song [2,3], Jia Qi [3], Wei Chen [3], Zhaoxiang Liu [3] and Ya Cheng [1,2,3,4,*]

[1] Engineering Research Center for Nanophotonics and Advanced Instrument, School of Physics and Electronic Science, East China Normal University, Shanghai 200241, China;
[2] State Key Laboratory of Precision Spectroscopy, School of Physics and Electronic Science, East China Normal University, Shanghai 200241, China;
[3] XXL—The Extreme Optoelectromechanics Laboratory, School of Physics and Electronic Science, East China Normal University, Shanghai 200241, China;
[4] State Key Laboratory of High Field Laser Physics, Shanghai Institute of Optics and Fine Mechanics, Chinese Academy of Sciences, Shanghai 201800, China
* Correspondence: jxu@phy.ecnu.edu.cn (J.X.); ya.cheng@siom.ac.cn (Y.C.)

**Abstract:** We demonstrate a hybrid laser microfabrication approach, which combines the technical merits of ultrafast laser-assisted chemical etching and carbon dioxide laser-induced in-situ melting, for centimeter-scale and bonding-free fabrication of 3D complex hollow microstructures in fused silica glass. With the developed approach, large-scale fused silica microfluidic chips with integrated 3D cascaded micromixing units can be reliably manufactured. High-performance on-chip mixing and continuous-flow photochemical synthesis under UV LEDs irradiation at ~280 nm were demonstrated using the manufactured chip, indicating a powerful capability for versatile fabrication of highly transparent all-glass microfluidic reactors for on-chip photochemical synthesis.

## 1. Introduction

Continuous-flow synthesis, in which a chemical reaction is performed in a flowing and confined environment (e.g., microchannels) other than conventional batch methods based on stirring, has attracted extensive attention in the past decade due to its unique advantages such as enhanced mass transfer, high throughput, low wastes, and high safety for manufacturing pharmaceuticals and fine chemicals [1-6]. In the continuous-flow synthesis, reliable fabrication of the microchannel reactors is of vital importance for practical industrial applications. As one of the most popular substrate materials, glass has many advantages such as high optical transparency, excellent chemical inertness, and low thermal expansion coefficients as compared to other materials (e.g., polymers, silicon, etc.) for microreactor applications [7-12]. Especially, fused silica has a broad transmission window from deep ultraviolet to infrared regions (200-3500 nm), which is an ideal substrate material for the versatile fabrication of photochemical microchannel reactors. In general, to fabricate a close glass-based microchannel reactor, the bonding of patterned glass workpieces usually is unavoidably employed using conventional fabrication methods such as lithographic methods and precision molding [12]. The validity and stability of bonding performance greatly affect the long-term operation of those microreactors, which increases the requirements of high-level surface quality and bonding environment of the glass workpieces. Currently, there is an increasing demand for the fabrication of microchannel reactors with three configurations (3D), which allows enhancement of the efficiency and the throughput of the device [13-16]. However, to form such a device with conventional methods will further increase the cost of the multiple-step bonding process. Therefore, the development of advanced fabrication technologies of 3D microchannel reactors in a bonding-free manner is highly desirable.

Ultrafast laser-assisted chemical etching of glass based on nonlinear multiphoton absorption provides a unique approach for the 3D fabrication of hollow glass microchannel structures [17-19]. In this technique, ultrafast laser direct writing in glass allows in-volume processing of 3D selectively modified regions under proper pulse energy. The modified regions can be further removed by etching solutions such as diluted HF solutions and hot KOH solutions to form the hollow channel structures [20-22]. In general, when the required lengths of microchannels were less than several centimeters, this technique can perform controllable fabrication of homogeneous microchannels without any bonding process. However, there are some technical barriers for fabricating long and large-volume microchannels with controllable feature sizes using this technique due to the inherent limitations of ultrafast laser-induced etching selectivity of glass [20-22]. Previously, to fabricate 3D centimeter-scale high-throughput micromixers in fused silica glass, a new bonding process was developed [23]. To beat the limitation of the etching selectivity, the introduction of extra-access ports was proposed by several groups before [24-27]. Recently, we have demonstrated that 3D all-glass microfluidic channels with arbitrary lengths can be fabricated using a hybrid laser microfabrication scheme based on a combination of ultrafast laser microfabrication and $CO_2$ laser irradiation of glass [28]. In the proposed hybrid scheme, ultrafast laser direct writing was employed to induce spatially selective modification inside the glass for subsequent chemical etching of hollow glass microstructures including 3D microchannels and throughout extra-access ports which were used for improving etching homogeneity of microchannels, and while defocusing $CO_2$ laser irradiation was employed to seal the etched extra-access ports on the glass surface to form a closed microfluidic structure with a few inlets and outlets. In this work, we demonstrate further progress on 3D large-scale fused silica microfluidic chips with integrated cascaded micromixing units that can be reliably manufactured using the improved hybrid scheme. With the improvement of $CO_2$ laser processing parameters, nearly perfect sealing of these ports has been realized in a controllable manner. Furthermore, high-performance on-chip mixing of dye solutions was demonstrated using the fabricated chip. And what's more, as for proof-of-concept demonstration, continuous-flow on-chip UV photochemical synthesis based on a photocycloaddition reaction has been realized using the same chip, indicating powerful capability for versatile manufacturing of highly transparent all-glass microfluidic systems for on-chip photochemical synthesis.

## 2. Materials and Methods

*Fabrication of 3D large-scale glass microchannels*

As illustrated in Figure 1a, the fabrication of 3D large-scale microchannels in fused silica consists of three main steps: (i) ultrafast laser direct writing, (ii) selective chemical etching, and (iii) defocusing $CO_2$ laser irradiation. For the ultrafast laser direct writing, fused silica glass plates (JGS2) with various sizes (e.g., 155 mm × 125 mm × 2 mm) were used for processing substrates and a laser amplifier system (Light Conversion Inc., Pharos 20W) with a central wavelength of 1030 nm, a pulse duration of 10 ps, and a repetition rate of 250 kHz was used for laser sources, respectively [29]. First, the glass plates were fixed on a 3D air-bearing stage system (Aerotech Inc., 3D X-Y-Z stage) with a positioning precision of 0.5 μm. Then, the laser beam was focused inside glass using a microscope objective (Mitutoyo, M Plan Apo NIR 10X) with a numerical aperture of 0.26 and a transmission rate of 80%. To create micropatterns with 3D configurations in glass, the modification was first performed from a line-by-line scanning in a single layer from the bottom of the patterns and then repeated upward layer by layer until the top of the patterns. The writing speed, line-by-line spacing, and layer-by-layer spacing were set at 65 mm/s, 30 µm, and 37.5 µm, respectively. As illustrated in Figure 1b, the laser-modified micropatterns include a 3D microchannel combined with an array of extra-access ports which connected the glass surface and the top of the channel. As previous works demonstrated [24-28], the introduction of extra-access ports provides a more homogeneous etching of the whole channel to beat the limitation of inherently etching selectivity of glass. For the selective chemical etching, a mixed aqueous solution with a concentration of 10 M KOH and 1 M NaOH was used. The laser-modified glass samples were immersed in the etching solution in an ultrasonic bath with a temperature of ~90 °C. After chemical etching, the 3D hollow glass channel structures were obtained and the etched extra-access ports were throughout from the glass surface to the inside of the 3D channel. For a glass plate with a size of 155 mm × 125 mm × 2 mm (see Fig. S1), the whole etching time was about 48 h. In addition, the distance between the top of the channel and the surface reduced from 1 mm to 700 µm, and while the height of the channel increased from 375 µm to 500 µm. For defocusing $CO_2$ laser irradiation, the etched glass plates were placed in a 2D stage (Coretech Inc., OneXY-500-500-AS-CMS1). The openings of the extra-access ports were irradiated by the $CO_2$ laser beam (Synrad, Inc., FSTi100SWC) with a wavelength of 10.6 µm and a repetition rate of 20 kHz one after another in a defocused manner using a ZnSe lens (Thorlabs, LA7028-E3) with a measured focal length of ~159 mm and finally sealed to form a close 3D large-scale microchannel with several inlets and outlets. With a defocusing distance of 21 mm, the irradiated diameter of the $CO_2$ laser beam on the glass surface was about 1 mm. To obtain the optimum sealing performance, different laser processing parameters such as $CO_2$ laser power and irradiation time were investigated.

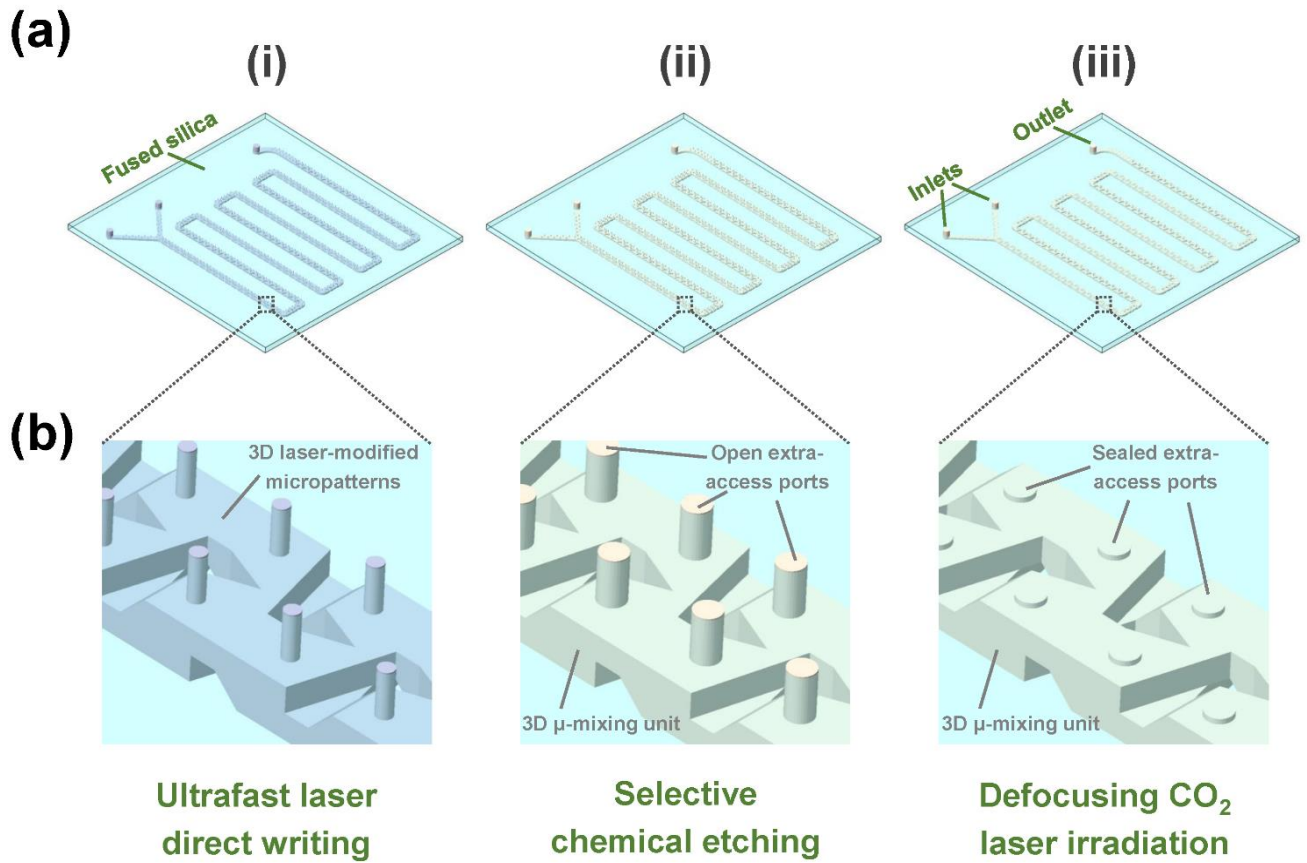

**Figure 1.** Schematic of the fabrication procedure for a 3D large-scale microchannel chip with two inlets and one outlet in fused silica. It consists of three main steps: (i) ultrafast laser direct writing of 3D modified micropatterns in glass, (ii) selective chemical etching of hollow glass microstructures including 3D microchannels, micromixing units, and open extra-access ports, and (iii) defocusing $CO_2$ laser irradiation for sealing the extra-access ports. (b) Close-up view of a 3D micromixing unit with several extra-access ports indicated by a dashed rectangle at each step in (a).

*Continuous-flow on-chip photochemical synthesis*

For on-chip photochemical reactions, the fabricated fused silica microfluidic chip with a size of 100 mm × 100 mm × 3 mm was closely arranged in front of a UV LED light source with a wavelength of 280 nm and an illuminated area of 90 mm × 75 mm. To perform high-performance on-chip mixing, 158 micromixing units with a cross-section of 2 mm × 1 mm were integrated inside the microfluidic chip. All reagents for on-chip continuous-flow synthesis were commercially available (Shanghai Aladdin Biochemical Technology Co., Ltd, Shanghai, China) and used without any purification. Reaction solutions were first prepared by dissolving maleimide (0.1 M) and 1-hexyne (0.15 M) into acetonitrile (MeCN), respectively. Then, both prepared solutions were simultaneously pumping into the 3D microfluidic chip through two inlets with the same flow rate ranging from 0.03 to 0.1 mL/min, respectively. The power of the UV LED light source was set at 160 mW. The products were collected from the outlet of the chip into an opaque glass bottle for subsequent nuclear magnetic resonance (NMR) measurements.

*Characterization*

The morphologies of the laser fabricated glass structures were recorded by a polarized optical microscope (Olympus, BX53). 1H NMR spectra of the on-chip continuous-flow synthesized products were recorded at 500 MHz (Bruker, AVANCE III HD500). For NMR characterization, the products were solved in the deuterated chloroform and tetramethylsilane (TMS) was used as an internal standard.

## 3. Results & Discussions

*Controllable sealing of extra-access ports on laser-fabricated glass microchannels*

Figure 2a shows cross-sectional optical micrographs of a laser-fabricated embedded microchannel structure with an ~240 μm diameter extra-access port (see Fig. S2) after different defocusing $CO_2$ laser irradiation conditions. By using defocusing $CO_2$ laser irradiation-induced glass in-situ melting, the extra-access port can be fully sealed to form a combined multilayer structure including a surface crater, a sealed layer, and a residual port from top to bottom as indicated in the top-left panel of Fig. 2a. The height of the sealed layer and residual port can be well controlled by tuning the laser power and irradiation time of the $CO_2$ laser. As shown in the top part of Fig.2a, when the laser power was set at 29.7 W, with the increase of irradiation time from 4 s to 10 s, the height of the sealed layer increased while the height of the residual port decreased, indicating the enhancement of time-dependent glass in-situ melting effects. Further increase of laser power to 31.3 W (the middle part of Fig.2a) and 32.6 W ((the bottom part of Fig.2a)), the height of the sealed layer continuously increased while the height of the residual port continuously decreased when the irradiation time was prolonged. Especially, when the irradiation time was 10 s at a laser power of 32.6 W, the height of the residual port was nearly eliminated in which there was almost no dead-end (nearly perfect sealing) in the fabricated microchannel structure.

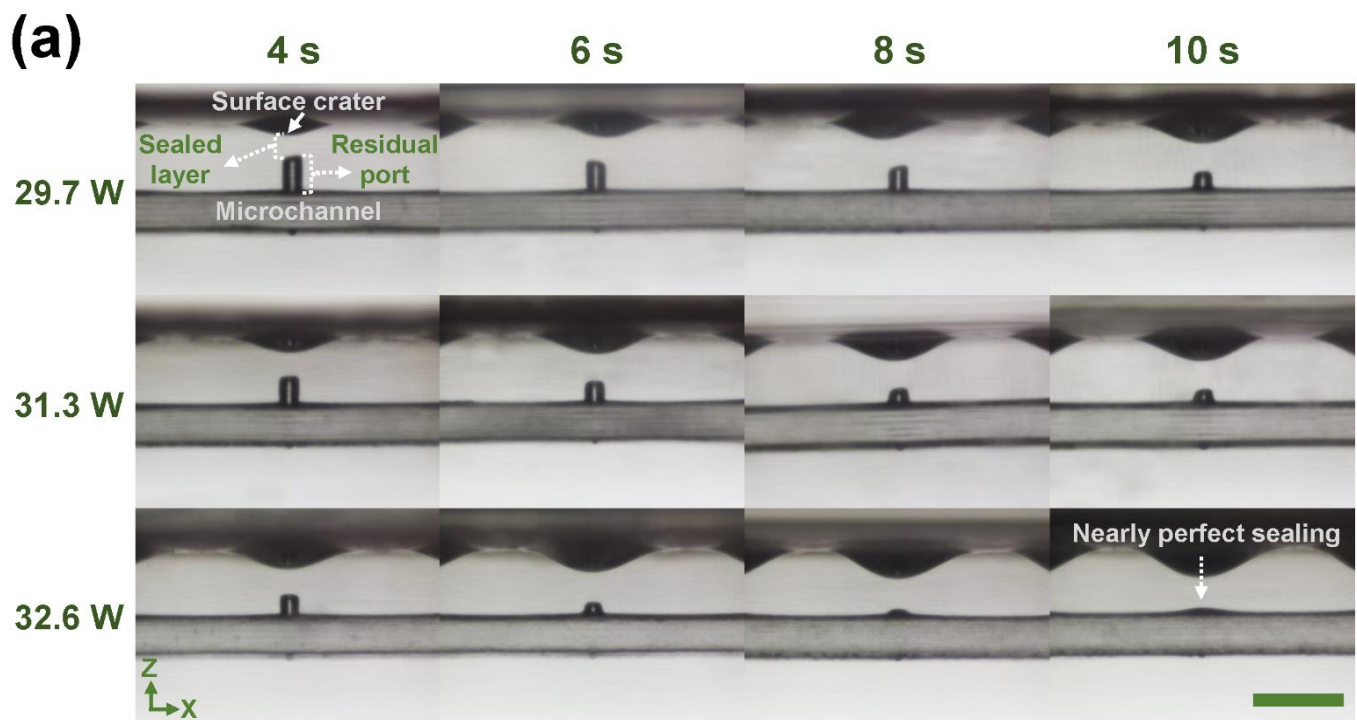

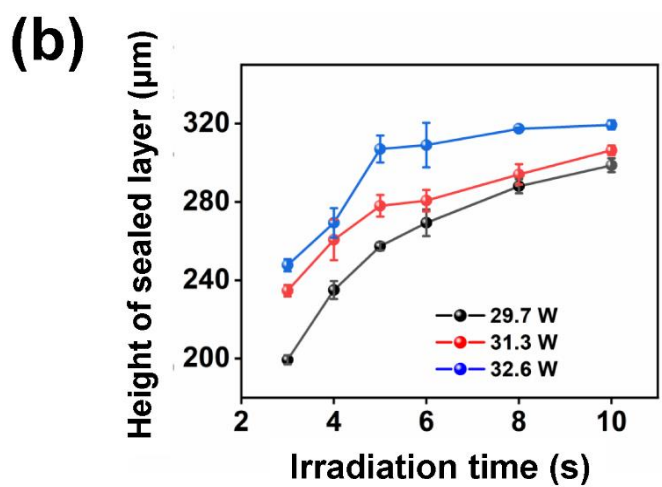
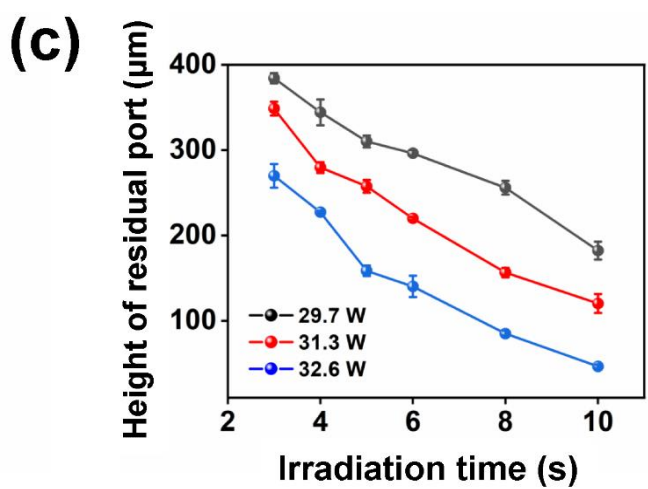

**Figure 2.** (a) Cross-sectional optical micrographs of a microchannel structure with an extra-access port after defocusing $CO_2$ laser irradiation at different laser powers (29.7, 31.3, and 32.6 W) and irradiation times (4, 6, 8, and 10 s). After $CO_2$ laser irradiation, a hollow extra-access port can be sealed to form a sealed layer while a surface crater is created and a part of the port residues depending on specific conditions. Scale bar represents 1 mm. (b) Height of sealed layer versus $CO_2$ laser irradiation time under different laser powers. (c) Height of residual port versus $CO_2$ laser irradiation time under different laser powers.

To further quantitively analyze the sealing performance of $CO_2$ laser processing of extra-access port, Figs. 2b and 2c plot the dependence of the height of sealed layer and residual port on irradiation time at different laser powers, respectively. As clearly presented in Figs. 2b and 2c, the extension of the irradiation time could enable the increase of the height of the sealed layer and the reduction of the height of the residual port, which exhibited the same tendency as the optical micrographs of Fig. 2a and Fig. S3a. Especially, in the case of 10 s irradiation at 32.6 W, a ~320 μm height sealed layer was obtained while the corresponded residual port was nearly eliminated. However, for a large-scale (e.g., centimeter scale) microfluidic chip that includes many extra-access ports to be sealed, reducing the sealing time for a single port may promote the processing efficiency for the whole chip. Therefore, there is a tradeoff between irradiation time and sealed performance. On the one hand, the port must be sealed with a proper sealed height (thickness) and a small residual volume in a certain amount of time for ensuring pressure-resistant microfluidic applications. On the other hand, the irradiation time must be kept in a short duration for rapid manufacturing.

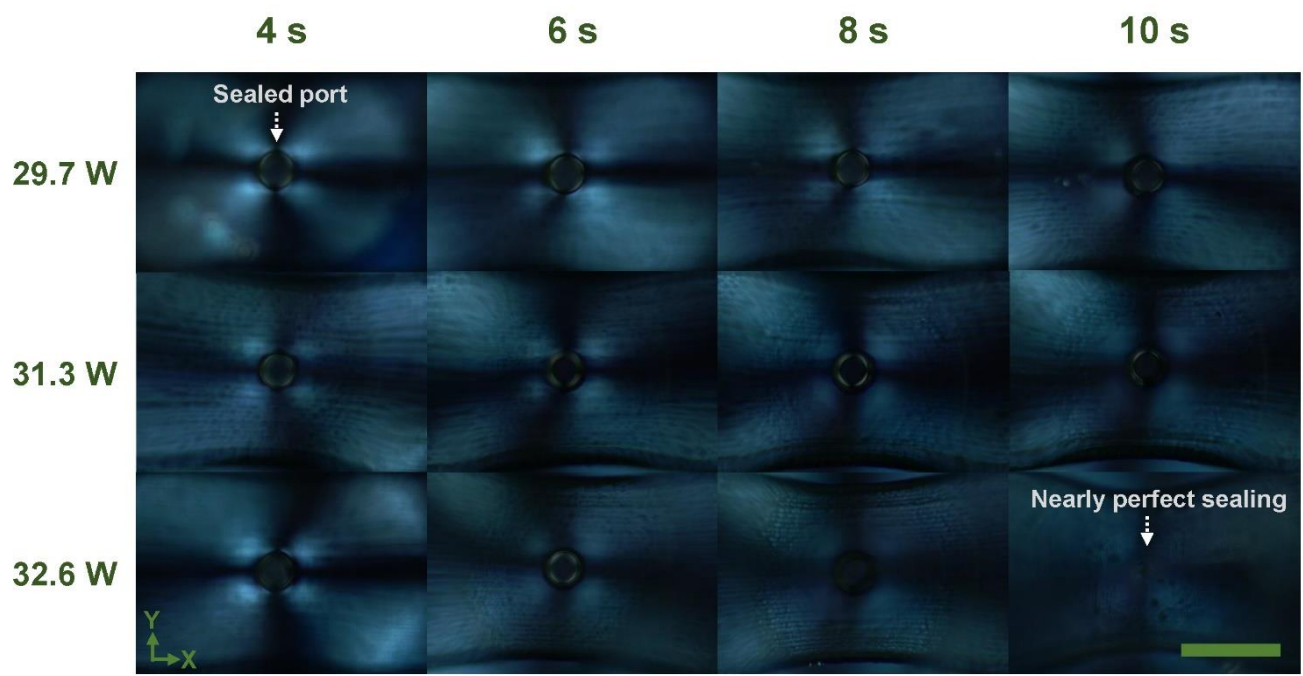

**Figure 3.** Front-view polarized optical micrographs of a microchannel structure with an extra-access port after defocusing $CO_2$ laser irradiation at different laser powers (29.7, 31.3, and 32.6 W) and irradiation times (4, 6, 8, 10 s). After $CO_2$ laser irradiation, the stress distribution of a sealed port exhibits differently depending on irradiation conditions. Scale bar represents 0.5 mm.

Besides the investigation of the size dependence of sealed ports on $CO_2$ laser processing parameters, polarized microscopic observation of the sealed ports has been also carried out. As shown in Fig. 3 and Fig. S3b, the polarized micrographs reveal the distribution of anisotropic birefringence around the ports after defocusing $CO_2$ laser irradiation. With the same laser power, the extension of irradiation time leads to the reduction of anisotropic birefringence, which promotes the homogeneity of the distribution of residual stress. Meanwhile, in the same irradiation time (6-10 s), the increase of laser power creates a similar tendency of stress control around the port. Especially, in the case of 10 s at 32.6 W, there is almost no anisotropic birefringence at the periphery of the port, indicating low residual stress in the condition of nearly perfect sealing as indicated in the bottom-right panel of Fig. 2a.

*Bonding-free fabrication of 3D large-scale micromixing fused silica chip*

To demonstrate the capability of the proposed approach for 3D large-scale fabrication of fused silica microfluidic chips, a T-shape centimeter-scale glass chip which included a string of 3D micromixing units was fabricated. Figures 4a and 4b show a schematic of fabricated 3D micromixing glass chip and photos of fabricated glass structure at each step, respectively. As illustrated in the inset of Fig.4a, the 3D micromixing unit is based on Baker's transformation, which consists of the functions of splitting, routing, and converging the fluids for high-efficiency mixing in 3D spaces [13, 14, 30]. Figure 4c shows cross-sectional optical micrographs of a 3D micromixing unit connected with an extra-access port at each fabrication step as described in Figure 4b. One can see in the left panel of Fig.4c that after ultrafast laser direct writing, 3D laser-written tracks which include the designable patterns of a 3D embedded microchannel beneath the glass surface and an extra-access port connected with the glass surface and the channel were created in a spatially selective manner. After selective chemical etching, the laser-modified micropatterns were transferred into a hollow glass structure including a 3D micromixing unit connected with an open port (see the middle panel of Fig.4c). Finally, the port can be sealed by defocusing $CO_2$ laser irradiation (4 s, 29.7 W) to form a 3D close micromixing unit, and then those mixing units can be cascaded to form a T-shape mixing glass chip with two inlets and one outlet in a robust and reproducible fashion.

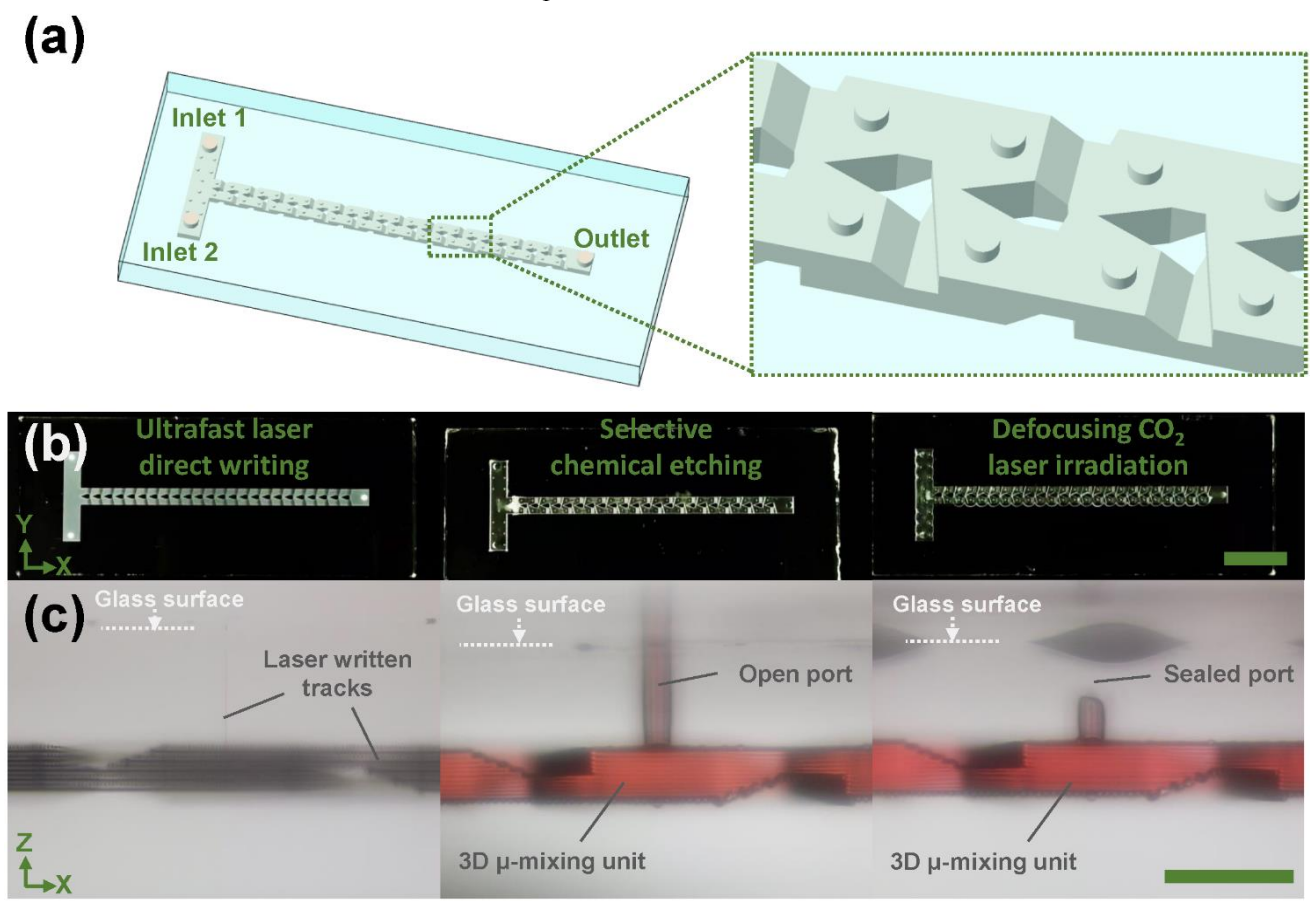

**Figure 4.** (a) Schematic of a centimeter-scale micromixing glass chip. Inset shows a close-up view of 3D micromixing units indicated by a dashed rectangle in the chip. (b) Photos of the glass chip at each fabrication step. Left: ultrafast laser direct writing; Middle: selective chemical etching; Right: Defocusing $CO_2$ laser irradiation. Scale bar represents 1 cm. (c) Cross-sectional optical micrographs of a 3D micromixing unit connected with an extra-access port at each fabrication step as described in (b). Left: laser written tracks; Middle: a 3D micromixing unit with an open port; Right: a 3D micromixing unit with a sealed port. The red color in (c) was the dye solution filled inside the channel, indicating the microfluidic function of the channel. Scale bar represents 1 mm.

*On-chip UV photochemical synthesis*

To verify the mixing performance of the fabricated fused silica chip, yellow and blue dye solutions were pumped into a fabricated fused silica chip with a size of 100 mm × 100 mm × 3 mm through two inlets shown in the left panel of Fig.5 a, respectively. With the embedded mixing units (the same as Fig. 4a) in the chip, the color of the mixing fluids quickly changed into green at the beginning of the entrance of the chip, indicating its high-performance mixing capability with a high throughput manner due to 3D configuration of the chip for manipulation of fluids, which can be also identified in the previous publication [23]. Moreover, by replacing the dye solutions with red and blue solutions, the purple solutions can be quickly obtained as shown in the right panel of Fig. 5a.

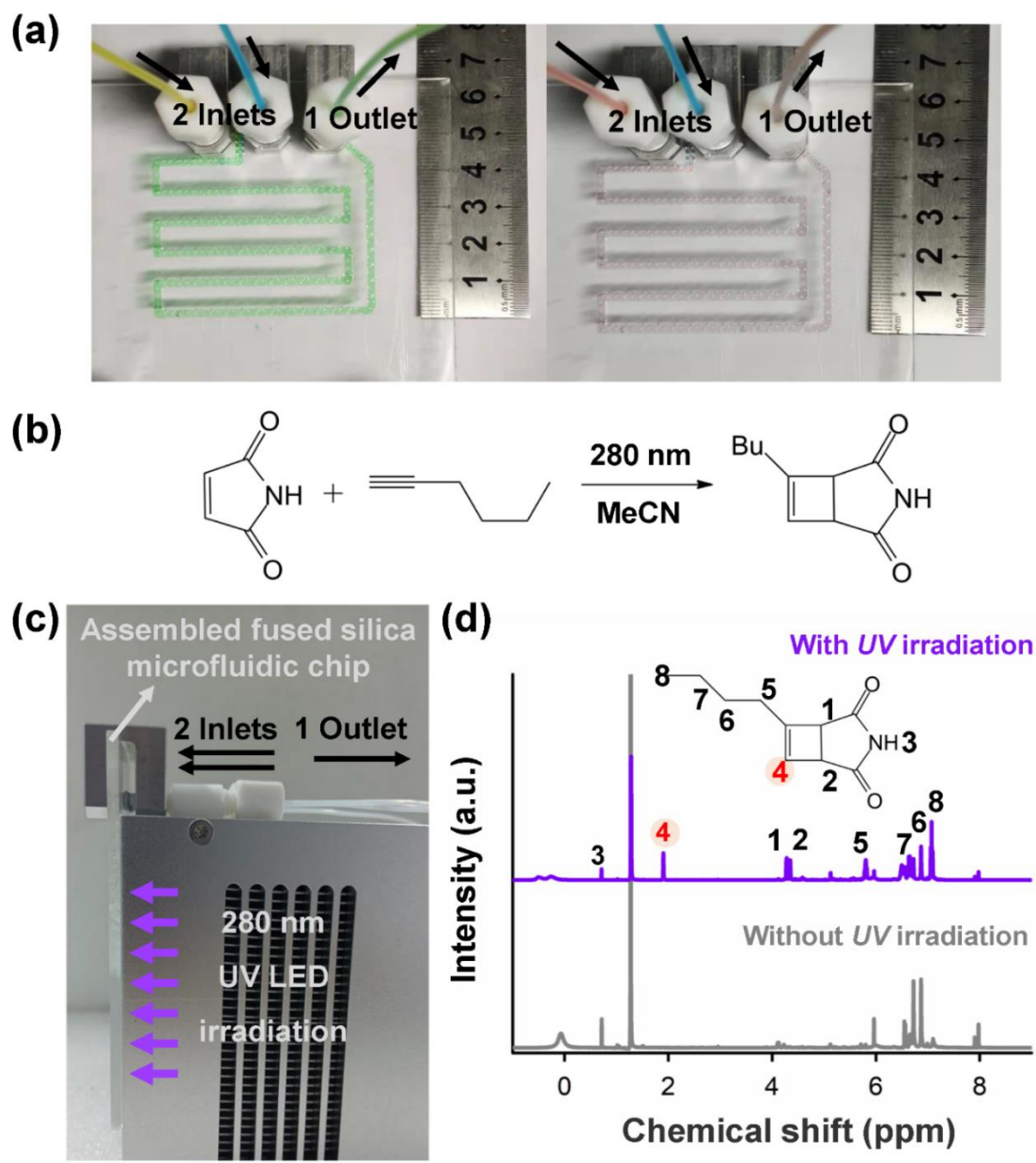

**Figure 5.** (a) Demonstration of high-efficiency mixing in a fused silica microfluidic chip with a size of 100 mm × 100 mm × 3 mm using different dye solutions. Left: yellow and blue dye solutions were pumped into two inlets with a flow rate of 20 ml/min, respectively; Right: red and blue dye solutions were pumped into two inlets with the same flow rate, respectively. (b) a [2+2] photocycloaddition reaction for continuous-flow photochemical synthesis. (c) Photo of an assembled fused silica microfluidic chip and a UV LED light source for on-chip photochemical reaction. (d) 1H NMR spectra of the synthesized product using continuous-flow on-chip synthesis with and without UV (280 nm) irradiation.

Regarding superior optical transmission properties of fused silica ranging from 200 nm to 3500 nm, on-chip UV photochemical synthesis using the same glass chip was performed. Cycloaddition reactions are reliable and effective in the synthesis of polycyclic compounds which are important for pharmaceutical engineering [31-33] and fine chemical engineering [34,35]. These reactions are usually driven by heat or light. Among the cycloaddition reactions, [2+2] photocycloaddition, which is driven by economical light irradiation, is widely used in the synthesis of natural products and medicines such as Taxol [36], one of the most commonly used anticancer agents, and ent-kaurane diterpenoids [37-39], an important group of natural products with antibacterial and anti-inflammatory activities. Therefore, as for proof-of-concept demonstration, a [2+2] photocycloaddition reaction (see Fig. 5b) for continuous-flow on-chip photochemical synthesis was investigated.

As illustrated in Fig. 5b, maleimide and 1-hexyne react under UV irradiation ($\lambda$ = ~280 nm). To perform highly effective UV irradiation, the assembled fused silica chip in Fig.5a was first closely attached to the illumination surface of a UV LED light source (280 nm) (see Fig. 5c) and performed the on-chip photochemical synthesis. The synthesized product was collected and dried under vacuum at 25 °C, then redissolved in deuterated chloroform for further NMR characterization. The 1H NMR spectra of the products which are obtained without and with UV irradiation at ~280 nm were shown in Figure 5d. A singlet peak which was remarked as No. 4 hydrogen represents the sign of the target molecule. Clearly, the reaction happens under UV irradiation since the singlet peak at 1.9 ppm emerged after irradiation, confirming the formation of the target molecule. In contrast, no reaction happens without UV irradiation. This result confirmed that a [2+2] photocycloaddition can work in the fabricated fused silica microfluidic chip under UV irradiation at ~280 nm.

## 4. Conclusions

Controllable fabrication of large-scale fused silica microfluidic chips with 3D configurations has been demonstrated using the hybrid strategy based on a combination of ultrafast laser microfabrication and $CO_2$ laser microprocessing. The introduction of extra-access ports connected with 3D microchannels allows the bonding-free manufacture of microfluidic chips with a workpiece size at a centimeter-scale level and a feature size at several hundreds of microns by beating the inherent limits of laser-induced etching selectivity of glass. Nearly perfect sealing of the ports has been stably obtained by optimizing the defocusing $CO_2$ laser irradiation. Further, fabrication of centimeter-scale microfluidic glass chips with cascaded micromixing units for high-efficiency and high-throughput mixing has been achieved, and the function of on-chip UV photochemical synthesis at ~280 nm has been verified. Regarding the unique capability of ultrafast laser microfabrication, the proposed approach provides a versatile route for the high-performance fabrication of large-scale fused silica microfluidic chips with monolithically multifunctional integration such as optofluidics and electrofluidics [20] for state-of-art sensing and detection, which will pave the way for rapid manufacturing of "all-in-one" glass-based microfluidic smart microsystems.

**Funding:** This research was funded by the National Natural Science Foundation of China (Grant Nos. 12174107, 61991444, 11933005, and 11734009), National Key R&D Program of China (Grant No. 2019YFA0705000), Science and Technology Commission of Shanghai Municipality (Grant No. 21DZ1101500).

**Conflicts of Interest:** The authors declare no conflict of interest.

**References**

1. Hartman, R. L.; McMullen, J. P.; Jensen K. F. Deciding Whether To Go with the Flow: Evaluating the Merits of Flow Reactors for Synthesis. *Angew. Chem. Int. Ed.* **2011**, 50, 7502-7519.
2. Elvira, K. S.; Solvas, X. C. i; Wootton, R. C. R.; de Mello, A. J. The past, present and potential for microfluidic reactor technology in chemical synthesis. *Nat. Chem.* **2013**, 5, 905-915.
3. Gutmann, B.; Cantillo, D.; Kappe C. O. Continuous-flow technology - a tool for the safe manufacturing of active pharmaceutical ingredients. *Angew. Chem. Int. Ed.* **2015**, 54, 6688-6728.
4. Adamo, A.; Beingessner, R. L.; Behnam, M.; Chen, J.; Jamison, T. F.; Jensen, K. F.; Monbaliu, J. C.; Myerson, A. S.; Revalor, E. M.; Snead, D. R.; Stelzer, T.; Weeranoppanant, N.; Wong, S. Y.; Zhang, P. On-demand continuous-flow production of pharmaceuticals in a compact, reconfigurable system. *Science* **2016**, 352, 61-67.
5. Bédard, A.-C.; Adamo, A.; Aroh, K. C.; Russell, M. G.; Bedermann, A. A.; Torosian, J.; Yue, B.; Jensen, K. F.; Jamison, T. F. Reconfigurable system for automated optimization of diverse chemical reactions. *Science* **2018**, 361, 1220-1225.
6. Buglioni, L.; Raymenants, F.; Slattery, A.; Zondag, S. D. A.; Noël, T. Technological Innovations in Photochemistry for Organic Synthesis: Flow Chemistry, High-Throughput Experimentation, Scale-up, and Photoelectrochemistry. *Chem. Rev.* **2022**, 122, 2752–2906.
7. Axinte, E. Glasses as engineering materials: A review. *Mater. Design* **2011**, 32, 1717-1732.
8. Ren, K.; Zhou, J.; Wu, H. Materials for Microfluidic Chip Fabrication. *Acc. Chem. Res.* **2013**, 46, 2396-2406.
9. Nge, P. N.; Rogers, C. I.; Woolley, A. T. Advances in Microfluidic Materials, Functions, Integration, and Applications. *Chem. Rev.* **2013**, 113, 2550-2583.
10. Hwang, J.; Cho, Y. H.; Park, M. S.; Kim, B. H. Microchannel Fabrication on Glass Materials for Microfluidic Devices. *Int. J. Precis. Eng. Manuf. Technol.* **2019**, 20, 479-495.
11. Gal-Or, E.; Gershoni, Y.; Scotti, G.; Nilsson, S. M. E.; Saarinen, J.; Jokinen, V.; Strachan, C. J.; Gennäs, G. B. A.; Yli-Kauhaluoma, J.; Kotiaho, T. Chemical analysis using 3D printed glass microfluidics. *Anal. Methods* **2019**, 11, 1802-1810.
12. Tang, T.; Yuan, Y.; Yalikun, Y.; Hosokawa, Y.; Li, M.; Tanaka, Y. Glass based micro total analysis systems: Materials, fabrication methods, and applications.Sensor. *Actuat. B - Chem.* **2021**, 339, 129859.
13. Yasui, T.; Omoto, Y.; Osato, K.; Kaji, N.; Suzuki, N.; Naito, T.; Watanabe, M.; Okamoto, Y.; Tokeshi, M.; Shamoto, E.; Baba, Y. Microfluidic baker's transformation device for three-dimensional rapid mixing. *Lab Chip* **2011**, 11, 3356-3360.
14. Liao, Y.; Song, J.; Li, E.; Luo, Y.; Shen, Y.; Chen, D.; Cheng, Y.; Xu, Z.; Sugioka, K.; Midorikawa, K. Rapid prototyping of three-dimensional microfluidic mixers in glass by femtosecond laser direct writing. *Lab Chip* **2012**, 12, 746-749.
15. Shallan, A. I.; Smejkal, P.; Corban, M.; Guijt, R. M.; Breadmore, M. C. Cost-Effective Three-Dimensional Printing of Visibly Transparent Microchips within Minutes. *Anal. Chem.* **2014**, 86, 3124-3130.
16. Enders, A.; Siller, I. G.; Urmann, K.; Hoffmann, M. R.; Bahnemann, J. 3D Printed Microfluidic Mixers - A Comparative Study on Mixing Unit Performances. *Small* **2019**, 15, 1804326.
17. Bellouard, Y.; Said, A.; Dugan, M.; Bado, P. Fabrication of high-aspect ratio, microfluidic channels and tunnels using femtosecond laser pulses and chemical etching. *Opt. Express* **2004**, 12, 2120-2129.
18. Kiyama, S.; Matsuo, S.; Hashimoto, S.; Morihira, Y. Examination of etching agent and etching mechanism on femtosecond laser microfabrication of channels inside vitreous silica substrates. *J. Phys. Chem. C* **2009**, 113, 11560-11566.
19. Gottmann, J.; Hermans, M.; Repiev, N.; Ortmann, J. Selective laser-induced etching of 3D precision quartz glass components for microfluidic applications - up-scaling of complexity and speed. *Micromachines* **2017**, 8, 110.
20. Sugioka, K.; Xu, J.; Wu, D.; Hanada, Y.; Wang, Z.; Cheng, Y.; Midorikawa, K. Femtosecond laser 3D micromachining: a powerful tool for the fabrication of microfluidic, optofluidic, and electrofluidic devices based on glass. *Lab Chip* **2014**, 14, 3447–3458.
21. Cheng, Y. Internal laser writing of high-aspect-ratio microfluidic structures in silicate glasses for lab-on-a-chip applications. *Micromachines* **2017**, 8, 59.
22. Ross, C. A.; Maclachlan, D. G.; Choudhury, D.; Thomson, R. R. Optimisation of ultrafast laser assisted etching in fused silica. *Opt. Express* **2018**, 26, 24343–24356.
23. Qi, J.; Li, W.; Chu, W.; Yu, J.; Wu, M.; Liang, Y.; Yin, D.; Wang, P.; Wang, Z.; Wang, M.; Cheng, Y. A Microfluidic Mixer of High Throughput Fabricated in Glass Using Femtosecond Laser Micromachining Combined with Glass Bonding. *Micromachines* **2020**, 11, 213.
24. He, F.; Lin, J.; Cheng, Y. Fabrication of hollow optical waveguides in fused silica by three-dimensional femtosecond laser micromachining. *Appl. Phys. B* **2011**, 105, 379-384.
25. Ho, S.; Herman, P. R.; Aitchison, J. S. Single- and multi-scan femtosecond laser writing for selective chemical etching of cross section patternable glass micro-channels. *Appl. Phys. A* **2012**, 106, 5-13.
26. He, S.; Chen, F.; Liu, K.; Yang, Q.; Liu, H.; Bian, H.; Meng, X.; Shan, C.; Si, J.; Zhao, Y.; Hou, X. Fabrication of three-dimensional helical microchannels with arbitrary length and uniform diameter inside fused silica. *Opt. Lett.* **2012**, 37, 18, 3825-3827.
27. Liu, Z.; Xu, J.; Lin, Z.; Qi, J.; Li, X.; Zhang, A.; Lin, J.; Chen, J.; Fang, Z.; Song, Y.; Chu, W.; Cheng, Y. Fabrication of single-mode circular optofluidic waveguides in fused silica using femtosecond laser microfabrication. *Opt. Laser Technol.* **2021**, 141, 107118.

28. Lin, Z.; Xu, J.; Song, Y.; Li, X.; Wang, P.; Chu, W.; Wang, Z.; Cheng, Y. Freeform Microfluidic Networks Encapsulated in Laser-Printed 3D Macroscale Glass Objects. *Adv. Mater. Technol.* **2020**, 5, 1900989.
29. Li, X.; Xu, J.; Lin, Z.; Qi, J.; Wang, P.; Chu, W.; Fang, Z.; Wang, Z.; Chai, Z.; Cheng, Y. Polarization-insensitive space-selective etching in fused silica induced by picosecond laser irradiation. *Appl. Surf. Sci.* **2019**, 485, 188-193.
30. Carrière, P. On a three-dimensional implementation of the baker's transformation. *Phys. Fluids* **2007**, 19, 118110.
31. Arena, G.; Chen, C. C.; Leonori, D.; Aggarwal, V. K. Concise Synthesis of (+)-allo-Kainic Acid via MgI2-Mediated Tandem Aziridine Ring Opening–Formal [3+2] Cycloaddition. *Org. Lett.* **2013**, 15, 4250-4253.
32. Qu, Z.-W.; Zhu, H.; Katsyuba, S. A.; Mamedova, V. L.; Mamedov, V. A.; Grimme, S. Acid-Catalyzed Rearrangements of 3-Aryloxirane-2-Carboxamides: Novel DFT Mechanistic Insights. *ChemistryOpen* **2020**, 9, 743-747.
33. Mashiko, T.; Shingai, Y.; Sakai, J.; Kamo, S.; Adachi, S.; Matsuzawa, A.; Sugita, K. Total Synthesis of Cochlearol B via Intra-molecular [2+2] Photocycloaddition. *Angew. Chem. Int. Ed.* **2021**, 60, 24484-24487.
34. Mackay, W. D.; Fistikci, M.; Carris, R. M.; Johnson, J. S. Lewis Acid Catalyzed (3 + 2)-Annulations of Donor–Acceptor Cyclopropanes and Ynamides. *Org. Lett.* **2014**, 16, 1626-1629.
35. Xu, C.-F.; Zheng, B.-H.; Suo, J.-J.; Ding, C.-H.; Hou, X.-L. Highly Diastereo- and Enantioselective Palladium-Catalyzed [3+2] Cycloaddition of Vinyl Aziridines and α,β-Unsaturated Ketones. *Angew. Chem. Int. Ed.* **2015**, 54, 1604-1607.
36. Schneider, F.; Samarin, K.; Zanella, S.; Gaich, T. Total synthesis of the complex taxane diterpene canataxpropellane. *Science* **2020**, 367, 676-681.
37. Cherney, E. C.; Green, J. C.; Baran, P. S. Synthesis of ent-Kaurane and Beyerane Diterpenoids by Controlled Fragmentations of Overbred Intermediates. *Angew. Chem. Int. Ed.* **2013**, 52, 9019-9022.
38. Chen, K.; Shi, Q.; Fujioka, T.; Zhang, D.-C.; Hu, C.-Q.; Jin, J.-Q.; Kilkuskie, R.-E.; Lee, K.-H. Anti-AIDS Agents, 4. Tripteri-fordin, a Novel Anti-HIV Principle from Tripterygium wilfordii: Isolation and Structural Elucidation. *J. Nat. Prod.* **1992**, 55, 88-92.
39. Kashiwada, Y.; Nishizawa, M.; Yamagishi, T.; Tanaka, T.; Nonaka, G.-i.; Cosentino, L. M.; Snider, J. V.; Lee, K.-H. Anti-AIDS Agents, 18. Sodium and Potassium Salts of Caffeic Acid Tetramers from Arnebia euchroma as Anti-HIV Agents. *J. Nat. Prod.* **1995**, 58, 392-400.